\RequirePackage{fix-cm}

\documentclass[twocolumn,epjc3]{svjour3}  
\smartqed  
\RequirePackage{graphicx}
\journalname{Eur. Phys. J. C}
\usepackage{amsmath}   
\usepackage{amssymb}   
\usepackage{graphicx}  
\usepackage{subcaption} 

\begin{document}

\title{Symmetry resolved entanglement in Schwinger pair production}


\author{
Wajid Joyia\thanksref{e1,addr1}
\and
S.M. Al-Marzoug\thanksref{addr2}
\and
Amaria Javed\thanksref{addr3}
}

\thankstext{e1}{e-mail: wajidjoyia@yahoo.com}

\institute{
Department of Physics, Quaid-i-Azam University, Islamabad, Pakistan\label{addr1}
\and
Physics Department and IRC-Advanced Quantum Computing, King Fahd University of Petroleum and Minerals, 31261 Dhahran, Saudi Arabia\label{addr2}
\and
Center for Quantum and Topological Systems, New York University Abu Dhabi, United Arab Emirates\label{addr3}
}

\authorrunning{Joyia et al.} 

\date{Received: date / Accepted: date}

\maketitle

\begin{abstract}
We study the charge resolved structure of the pairwise mode entanglement produced by the Schwinger effect. Using the $U(1)$ electric charge of quantum electrodynamics, we decompose the entanglement of the reduced density matrices of momentum modes into individual charge sectors, applying
the symmetry resolved entanglement entropy to the pure bipartition reductions and the charge imbalance resolved negativity to the mixed particle--antiparticle pair. For a constant electric field we obtain closed form results for every mode pairing, with all field dependence
controlled by the single pair production probability
$|\beta|^2=e^{-\pi\mu}$. For the particle--particle and
antiparticle--antiparticle pairs the entanglement entropy within each charge sector is field independent, so that the field acts only by redistributing statistical weight among the sectors; for the mixed particle--antiparticle pair the populated sector additionally carries a field dependent entropy. The vacuum created pair is the richest case: its
resolved negativity is distributed over three imbalance sectors and is transported from one sector to the next as the field increases, the sectors contributing equally only at the critical field. The entanglement across a
fixed momentum bipartition, by contrast, is frozen at its initial value because pair creation conserves the sector charge. Symmetry resolution thus refines the total entanglement measures into a sector resolved picture,
exposing where in charge space the entanglement of the Schwinger mechanism resides and how the electric field moves it—information beyond that contained in the total measures or in particle number distributions.
\end{abstract}
\section{Introduction}
\label{intro}

Building on earlier work by Sauter~\cite{Sauter1931} and Heisenberg and Euler~\cite{HeisenbergEuler1936}, Schwinger showed that a strong electric field renders the quantum vacuum unstable against the creation of electron--positron pairs~\cite{Schwinger1951}. This Schwinger effect is a nonperturbative prediction of quantum electrodynamics whose rate, $\sim\exp[-\pi m^2/(eE)]$, has so far precluded direct observation; it remains a central goal in strong laser and heavy ion physics~\cite{DiPiazza2012,Cooper1993}, and numerous schemes have been proposed to enhance its detectability~\cite{Gies2005,Dunne2009,%
Schutzhold2008,DumluDunne2010,Hebenstreit2011}. As a quantum process in a classical background that converts vacuum fluctuations into real particles, the Schwinger effect is formally analogous to the Unruh effect, to Hawking radiation, and to cosmological particle production~\cite{Brout1995,Martin2008}, all of which are governed by Bogoliubov transformations relating the ``in'' and ``out'' mode
bases~\cite{Kim2008}.

In parallel, quantum information science has provided new tools for characterizing quantum correlations in field theory. A natural way to define entanglement in a field theoretic setting is to treat the modes, labelled by coordinates or momenta, as the subsystems~\cite{Shi2004}. This viewpoint has been applied extensively to relativistic and gravitational settings, including quantum fields near black holes~\cite{FuentesSchuller2005,%
MartinMartinez2010,Dai2015,Dai2016,joyia2025,joyia2026} and in expanding universes~\cite{Ball2006,Fuentes2010,Moradi2014}. The Schwinger effect has likewise been examined from this perspective: the entanglement between a single Dirac mode and the rest of the system was computed in Ref.~\cite{EbadiMirza2014}, and the pairwise mode entanglement generated and redistributed by the field was studied in detail in Ref.~\cite{LiDaiShi2016} using the mutual information and the logarithmic negativity. More recent work has explored the quantum metrology of the Schwinger effect~\cite{Fan2024} and the real time growth of entanglement in the massive Schwinger model as a proxy for jet production and hadronization~\cite{Florio2023,Florio2024}.

Whereas these studies characterize the total entanglement between modes, a finer description is available when the theory possesses a global symmetry. Quantum electrodynamics conserves the $U(1)$ electric charge, and the reduced density matrices of the created pairs therefore commute with a local charge operator. This makes it possible to decompose the entanglement into contributions from individual charge sectors, a program known as symmetry resolved entanglement~\cite{Goldstein2018,Laflorencie2014,Xavier2018}. The total entanglement entropy then splits
into a configurational part and a number (fluctuation)
part~\cite{Wiseman2003,Bonsignori2019}, while for mixed states the symmetry resolved negativity admits an analogous decomposition into charge imbalance sectors~\cite{Cornfeld2018}; a careful treatment of the latter for fermions requires the fermionic partial transpose~\cite{Shapourian2017,Eisler2015}. Symmetry resolved entanglement has become a standard diagnostic across condensed matter and field theory contexts, and charge resolved entanglement spectra have recently been computed even in the massive Schwinger model~\cite{Florio2024}. Its application to the Schwinger production of particle--antiparticle pairs in a prescribed electric field has, however, not been carried out, and it is the subject of the present work.

In this paper we resolve, sector by sector, the pairwise mode entanglement of the Schwinger effect. Starting from the reduced density matrices of Ref.~\cite{LiDaiShi2016}, we apply the symmetry resolved entanglement entropy to the particle--particle and antiparticle--antiparticle mode pairs, and the symmetry resolved  negativity to the particle--antiparticle pairs, choosing in each case the measure appropriate to whether the reduced state is a pure bi-partition reduction or
a mixed state. Working with a constant electric field, we obtain
closed form results for every mode pairing and find that the entanglement within each charge sector is field independent, so that the field acts purely by redistributing statistical weight among the sectors, with all dependence controlled by the pair production probability $|\beta|^2=e^{-\pi\mu}$. In particular, the negativity of the vacuum created pair is genuinely distributed over several charge sectors and is transported between them as the field is varied, while the entanglement across a fixed momentum bi-partition is frozen at its initial
value by charge conservation. The paper is organized as follows. In Sec.2 we review the mode structure and the reduced density matrices and define the symmetry resolved quantities. Sections.3 present the resolved entropy and negativity for the
various mode pairings, and Sec.4 summarizes our
findings and discusses extensions to the pulsed field and to related particle creation processes.
\section{Formalism}
\label{sec:setup}

\subsection{Dirac field and Bogoliubov transformation}

We consider a Dirac field $\Psi(x)$ of mass $m$ and charge $e$ in a
spatially uniform electric field $\mathbf{E}=E_0\hat{\mathbf z}$, described
by the potential $A_\mu=(0,0,0,-E_0 t)$. The field obeys the Dirac
equation
\begin{equation}
  \bigl(i\gamma^\mu\partial_\mu-e\gamma^\mu A_\mu-m\bigr)\Psi(x)=0 .
  \label{eq:dirac}
\end{equation}
Let $\{u^{\rm in},v^{\rm in}\}$ and $\{u^{\rm out},v^{\rm out}\}$ be two
complete sets of mode solutions of Eq.~\eqref{eq:dirac}, defined
asymptotically at $t\to-\infty$ (``in'') and $t\to+\infty$ (``out'')
respectively. The field operator admits the two equivalent expansions
\begin{equation}
  \begin{aligned}
  \Psi(x)&=\sum_{\sigma,\mathbf{k}}
  \Bigl[a^{\rm in}_{\sigma,\mathbf{k}}u^{\rm in}_{\sigma,\mathbf{k}}(x)
       +b^{\rm in\dagger}_{\sigma,\mathbf{k}}v^{\rm in}_{\sigma,\mathbf{k}}(x)\Bigr]\\
  &=\sum_{\sigma,\mathbf{k}}
  \Bigl[a^{\rm out}_{\sigma,\mathbf{k}}u^{\rm out}_{\sigma,\mathbf{k}}(x)
       +b^{\rm out\dagger}_{\sigma,\mathbf{k}}v^{\rm out}_{\sigma,\mathbf{k}}(x)\Bigr],
  \end{aligned}
  \label{eq:expansion}
\end{equation}
where $a_{\sigma,\mathbf{k}}$ and $b_{\sigma,\mathbf{k}}$ annihilate
particles and antiparticles of spin $\sigma=\uparrow,\downarrow$ and
momentum $\mathbf{k}$. The two sets of operators are connected by a
Bogoliubov transformation~\cite{Kim2008,LiDaiShi2016},
\begin{align}
  a^{\rm in}_{\sigma,\mathbf{k}}
    &=\alpha_{\mathbf{k}}^{*}\,a^{\rm out}_{\sigma,\mathbf{k}}
     +\beta_{\mathbf{k}}^{*}\,b^{\rm out\dagger}_{-\sigma,-\mathbf{k}},
     \label{eq:bogo_a}\\
  b^{\rm in}_{\sigma,\mathbf{k}}
    &=\alpha_{\mathbf{k}}^{*}\,b^{\rm out}_{\sigma,\mathbf{k}}
     -\beta_{\mathbf{k}}^{*}\,a^{\rm out\dagger}_{-\sigma,-\mathbf{k}},
     \label{eq:bogo_b}
\end{align}
with the coefficients
\begin{equation}
  \alpha_{\mathbf{k}}=\sqrt{\tfrac{\mu}{\pi}}\,
    \Gamma\!\bigl(\tfrac{i\mu}{2}\bigr)\sinh\!\bigl(\tfrac{\pi\mu}{2}\bigr)
    e^{-\pi\mu/4}, \qquad
  \beta_{\mathbf{k}}=e^{-\pi\mu/2},
  \label{eq:coeffs}
\end{equation}
so that
\begin{equation}
  \begin{aligned}
  &|\alpha_{\mathbf{k}}|^2=1-e^{-\pi\mu},\qquad
   |\beta_{\mathbf{k}}|^2=e^{-\pi\mu},\\
  &\mu=\frac{m^2+k_\perp^2}{eE_0},\qquad
   k_\perp^2=k_x^2+k_y^2 .
  \end{aligned}
  \label{eq:mu}
\end{equation}
The quantity $|\beta_{\mathbf{k}}|^2=e^{-\pi\mu}$ is the probability of
creating a pair in the mode $\mathbf{k}$; it increases monotonically with
$E_0$ and controls all of the results below. Because pairs are created in
modes of opposite momenta, it is natural to work within the sector spanned
by the modes $\mathbf{k}$ and $-\mathbf{k}$. Each momentum mode is treated
as a subsystem~\cite{Shi2004}, i.e.\ a four-dimensional Hilbert space with
occupation basis $\{|0\rangle,|\!\uparrow\rangle,|\!\downarrow\rangle,
|\!\uparrow\downarrow\rangle\}$, so that spin correlations between distinct
momentum modes are retained.

\subsection{Pair creation from the vacuum}

The in-vacuum is annihilated by the in operators,
$a^{\rm in}_{\sigma,\mathbf{k}}|0\rangle^{\rm in}=
 b^{\rm in}_{\sigma,\mathbf{k}}|0\rangle^{\rm in}=0$. Imposing
$a^{\rm in}_{\sigma,\mathbf{k}}|0_{\mathbf{k}},0_{-\mathbf{k}}\rangle^{\rm in}=0$
with Eq.~\eqref{eq:bogo_a} determines the in-vacuum as a superposition of
out-mode occupations~\cite{LiDaiShi2016},
\begin{equation}
  \begin{aligned}
  |0_{\mathbf{k}},0_{-\mathbf{k}}\rangle^{\rm in}
  ={}&|\alpha_{\mathbf{k}}|^2\,|0_{\mathbf{k}},0_{-\mathbf{k}}\rangle^{\rm out}
  -\alpha_{\mathbf{k}}\beta_{\mathbf{k}}^{*}
     |\!\uparrow_{\mathbf{k}},\downarrow_{-\mathbf{k}}\rangle^{\rm out}\\
  &-\alpha_{\mathbf{k}}\beta_{\mathbf{k}}^{*}
     |\!\downarrow_{\mathbf{k}},\uparrow_{-\mathbf{k}}\rangle^{\rm out}
  +\frac{\alpha_{\mathbf{k}}\beta_{\mathbf{k}}^{*2}}{\alpha_{\mathbf{k}}^{*}}
     |\!\uparrow\downarrow_{\mathbf{k}},\uparrow\downarrow_{-\mathbf{k}}\rangle^{\rm out}.
  \end{aligned}
  \label{eq:invac}
\end{equation}
The density matrix
$\rho_{\mathbf{k},-\mathbf{k}}=
 |0_{\mathbf{k}},0_{-\mathbf{k}}\rangle^{\rm in}\langle
 0_{\mathbf{k}},0_{-\mathbf{k}}|$ is a pure entangled state of the particle
mode $\mathbf{k}$ and the antiparticle mode $-\mathbf{k}$; its reduced state
$\rho_{\mathbf{k}}=\mathrm{Tr}_{-\mathbf{k}}\,\rho_{\mathbf{k},-\mathbf{k}}$
is diagonal, with eigenvalues $|\alpha|^4,|\alpha|^2|\beta|^2,
|\alpha|^2|\beta|^2,|\beta|^4$, which we resolve by the mode charge
$Q=n_{\mathbf{k}}$ in Sec.3.4.

\subsection{Action of the field on an initially entangled pair}

To study how the field acts on pre-existing correlations, we take as
initial state two particles of momenta $\mathbf{p}$ and $\mathbf{q}$ in a
spin-entangled state~\cite{LiDaiShi2016},
\begin{equation}
  \varepsilon\,|\!\uparrow\rangle_{\mathbf{p}}|\!\downarrow\rangle_{\mathbf{q}}
  +\sqrt{1-\varepsilon^{2}}\,
  |\!\downarrow\rangle_{\mathbf{p}}|\!\uparrow\rangle_{\mathbf{q}},
  \label{eq:initial}
\end{equation}
where $\varepsilon$ sets the initial entanglement. In terms of in modes,
\begin{equation}
  \begin{aligned}
  |\Phi_{\mathbf{p},\mathbf{q},-\mathbf{p},-\mathbf{q}}\rangle^{\rm in}
  ={}&\bigl(\varepsilon\,|\!\uparrow_{\mathbf{p}},\downarrow_{\mathbf{q}}\rangle^{\rm in}
   +\sqrt{1-\varepsilon^{2}}\,
    |\!\downarrow_{\mathbf{p}},\uparrow_{\mathbf{q}}\rangle^{\rm in}\bigr)\\
  &\times|0_{-\mathbf{p}},0_{-\mathbf{q}}\rangle^{\rm in}.
  \end{aligned}
  \label{eq:Phiin}
\end{equation}
Using $a^{\rm in\dagger}_{\sigma,\mathbf{k}}=
\alpha_{\mathbf{k}}a^{\rm out\dagger}_{\sigma,\mathbf{k}}
+\beta_{\mathbf{k}}b^{\rm out}_{-\sigma,-\mathbf{k}}$ acting on the
in-vacuum yields the single-particle relations, valid for any mode
$\mathbf{k}$,
\begin{align}
  |\!\uparrow_{\mathbf{k}},0_{-\mathbf{k}}\rangle^{\rm in}
   &=\alpha_{\mathbf{k}}|\!\uparrow_{\mathbf{k}},0_{-\mathbf{k}}\rangle^{\rm out}
    -\frac{\alpha_{\mathbf{k}}\beta_{\mathbf{k}}^{*}}{\alpha_{\mathbf{k}}^{*}}
     |\!\uparrow\downarrow_{\mathbf{k}},\uparrow_{-\mathbf{k}}\rangle^{\rm out},
     \label{eq:up}\\
  |\!\downarrow_{\mathbf{k}},0_{-\mathbf{k}}\rangle^{\rm in}
   &=\alpha_{\mathbf{k}}|\!\downarrow_{\mathbf{k}},0_{-\mathbf{k}}\rangle^{\rm out}
    +\frac{\alpha_{\mathbf{k}}\beta_{\mathbf{k}}^{*}}{\alpha_{\mathbf{k}}^{*}}
     |\!\uparrow\downarrow_{\mathbf{k}},\downarrow_{-\mathbf{k}}\rangle^{\rm out}.
     \label{eq:down}
\end{align}
Applying Eqs.~\eqref{eq:up}--\eqref{eq:down} with $\mathbf{k}=\mathbf{p}$
and $\mathbf{k}=\mathbf{q}$ in Eq.~\eqref{eq:Phiin} expands
$|\Phi\rangle^{\rm in}$ over the sixteen out-mode configurations of the
four modes $(\mathbf{p},-\mathbf{p},\mathbf{q},-\mathbf{q})$. The reduced density
matrices studied below are obtained by tracing this pure state over the
appropriate complementary modes,
\begin{equation}
  \rho_{X}=\mathrm{Tr}_{\bar X}
  \bigl(|\Phi\rangle^{\rm in}\langle\Phi|\bigr),
  \label{eq:reduce}
\end{equation}
where $X$ denotes the retained mode pair and $\bar X$ its complement. For
example, tracing over $(-\mathbf{p},-\mathbf{q})$ gives the
particle--particle matrix $\rho_{\mathbf{p},\mathbf{q}}$; tracing over
$(-\mathbf{p},\mathbf{q})$ gives the particle--antiparticle matrix
$\rho_{\mathbf{p},-\mathbf{q}}$; and tracing over $(\mathbf{q},-\mathbf{q})$
gives the same-momentum partner matrix $\rho_{\mathbf{p},-\mathbf{p}}$. The
orthonormality of the out-mode occupation states makes each partial trace
elementary: only terms sharing the same occupation of the traced modes
interfere. These reduced matrices reproduce those of
Ref.~\cite{LiDaiShi2016} and form the input to the symmetry resolved
analysis of the following sections.

\subsection{Symmetry resolved entanglement measures}
\label{sec:sre_defs}

Quantum electrodynamics conserves the $U(1)$ electric charge, so each
reduced state $\rho_X$ commutes with the physical charge operator of its
retained modes, $\hat{Q}_X=\sum_{\text{modes}}(\hat{n}_{\rm part}
-\hat{n}_{\rm antipart})$, i.e.\ the number of particles minus the number
of antiparticles; $[\rho_X,\hat{Q}_X]=0$. We resolve each entropy with
respect to this physical charge, which we denote $Q$. For an all particle (or all antiparticle) subsystem $Q$ coincides with the total occupation,
whereas for a mixed particle--antiparticle subsystem the two differ, and it
is the physical charge $Q$ that we use throughout. The reduced state is
then block-diagonal in the eigenvalue $Q$,
\begin{equation}
  \rho_X=\bigoplus_Q p_Q\,\rho_X^{(Q)},
  \
  p_Q=\mathrm{Tr}\!\bigl(\Pi_Q\,\rho_X\bigr),
  \
  \rho_X^{(Q)}=\frac{\Pi_Q\,\rho_X\,\Pi_Q}{p_Q},
  \label{eq:blockdiag}
\end{equation}
where $\Pi_Q$ projects onto the charge-$Q$ sector, $\rho_X^{(Q)}$ is the
normalized state within it, and $p_Q\ge0$ with $\sum_Q p_Q=1$ is the
probability of finding charge $Q$ in the subsystem.

\paragraph{Symmetry resolved entanglement entropy.}
When $\rho_X$ is the reduced state of a pure global state across a
bipartition, its von Neumann entropy
$S(\rho_X)=-\mathrm{Tr}(\rho_X\log_2\rho_X)$ is the entanglement entropy of
that cut. The symmetry-resolved entanglement entropy is the entropy within
a fixed charge sector~\cite{Goldstein2018},
\begin{equation}
  S_Q\equiv S\bigl(\rho_X^{(Q)}\bigr)
  =-\mathrm{Tr}\!\bigl(\rho_X^{(Q)}\log_2\rho_X^{(Q)}\bigr),
  \label{eq:Sq}
\end{equation}
and the total entropy decomposes exactly as~\cite{Wiseman2003,Laflorencie2014}
\begin{equation}
  S(\rho_X)=\underbrace{\sum_Q p_Q\,S_Q}_{S_{\rm config}}
           +\underbrace{\Bigl(-\sum_Q p_Q\log_2 p_Q\Bigr)}_{S_{\rm num}} .
  \label{eq:decomp}
\end{equation}
The configurational entropy $S_{\rm config}$ measures the entanglement
carried within the sectors and is the part accessible under the charge
superselection rule, while the number (fluctuation) entropy $S_{\rm num}$
measures the entropy of the charge distribution itself~\cite{Bonsignori2019}.
The probabilities $p_Q$ are computed directly from the eigenvalues of
$\rho_X$ grouped by sector, and the $S_Q$ from the eigenvalues of each
normalized block $\rho_X^{(Q)}$.

\paragraph{Symmetry resolved negativity.}
For the mixed particle--antiparticle reduced states the von Neumann
entropy is not an entanglement measure, and we use the logarithmic
negativity~\cite{Vidal2002}. For a bipartition $X=A\cup B$,
\begin{equation}
  \mathcal{N}(\rho_X)=\sum_{\lambda_i<0}|\lambda_i|,
  \
  N(\rho_X)=\log_2\bigl\|\rho_X^{T_A}\bigr\|_1
          =\log_2\!\bigl(1+2\mathcal{N}\bigr),
  \label{eq:neg}
\end{equation}
where $\lambda_i$ are the eigenvalues of the partial transpose
$\rho_X^{T_A}$ and $\|\cdot\|_1$ is the trace norm. Here the resolving
charge is not the subsystem charge $\hat{Q}_A$ but the charge imbalance
$\hat{q}=\hat{Q}_A-\hat{Q}_B$, which is the operator that commutes with the
partial transpose $\rho_X^{T_A}$~\cite{Cornfeld2018}; we denote its
eigenvalue $q$ to distinguish it from the entropy charge $Q$. The partial
transpose then block-diagonalizes into imbalance sectors, and the
symmetry resolved negativity is obtained by summing the moduli of the
negative eigenvalues within each sector,
\begin{equation}
  \mathcal{N}(q)=\!\!\sum_{\lambda_i<0,\ i\in q}\!\!|\lambda_i|,
  \qquad
  \sum_q \mathcal{N}(q)=\mathcal{N}.
  \label{eq:Nq}
\end{equation}
Unlike the entropy, the negativity carries no sector probabilities, since
$\rho_X^{T_A}$ is not a physical state; the resolution consists solely of
the per sector negativities $\mathcal{N}(q)$. We note that for particle--antiparticle modes a rigorous definition may
require the fermionic partial transpose~\cite{Shapourian2017,Eisler2015};
for the vacuum pair we verify in Appendix~A that the ordinary and fermionic
prescriptions coincide, and we comment on the mixed cross terms and partner
matrices where they arise.

All of the reduced matrices in this work are finite-dimensional, so the
spectra entering Eqs.~\eqref{eq:Sq}--\eqref{eq:Nq} are obtained in closed
form, and the resolved quantities follow directly from the Bogoliubov
coefficients~\eqref{eq:mu}.

\subsection{Choice of entanglement measure for each mode pairing}

Which resolved quantity is appropriate depends on the mode pairing. For
the particle--particle state $\rho_{\mathbf{p},\mathbf{q}}$---a reduced
density matrix of the pure four-mode state across the bipartition
$(\mathbf{p},\mathbf{q})\,|\,(-\mathbf{p},-\mathbf{q})$---the von Neumann
entropy is a genuine entanglement entropy, and its symmetry resolved version is both well defined and structurally rich: the entanglement
distributes over the charge sectors
$Q=n_{\mathbf{p}}+n_{\mathbf{q}}=2,3,4$. We therefore characterize the
particle--particle and antiparticle--antiparticle pairings, together with
the particle--antiparticle reductions of the pure four-mode state, by their
symmetry-resolved entanglement entropy.

For the pairwise entanglement of a mixed particle--antiparticle state
between its two modes, the von Neumann entropy is not an entanglement
measure, and we use the symmetry resolved negativity. This is the case for
the vacuum pair $\rho_{\mathbf{k},-\mathbf{k}}$, whose resolved negativity
spreads over several imbalance sectors whose weights are redistributed by
the field structure invisible to the total negativity. In short, the
measure is fixed by whether one is quantifying the entropy of a
pure-bipartition reduction (SREE) or the entanglement of a mixed pairwise
state (negativity), and we resolve the former by the physical charge $Q$
and the latter by the imbalance charge $q$.
\section{Results}

Throughout, we resolve each reduced state by the physical $U(1)$
charge of its retained modes, the number of particles minus the number of
antiparticles and denote it $Q$. For the entropy this is the subsystem
charge $Q_A$; for the negativity it is the charge imbalance
$Q_A-Q_B$, which we denote $q$ to distinguish it. For an all-particle
subsystem the physical charge coincides with the total occupation, whereas
for a mixed particle--antiparticle subsystem the two differ, and it is the
physical charge that we use.

\subsection{Particle--particle modes}
\begin{figure}[t]
  \centering
  \includegraphics[width=0.48\textwidth]{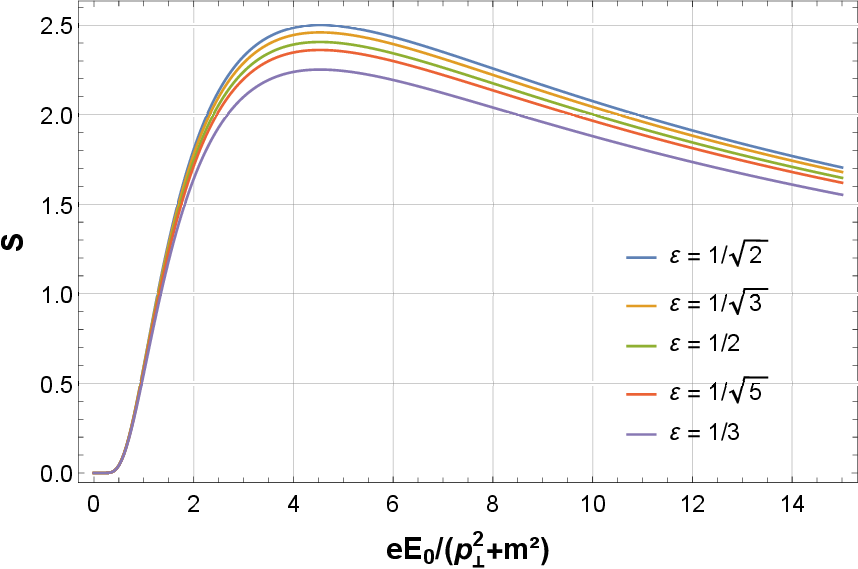}\hfill
  \includegraphics[width=0.48\textwidth]{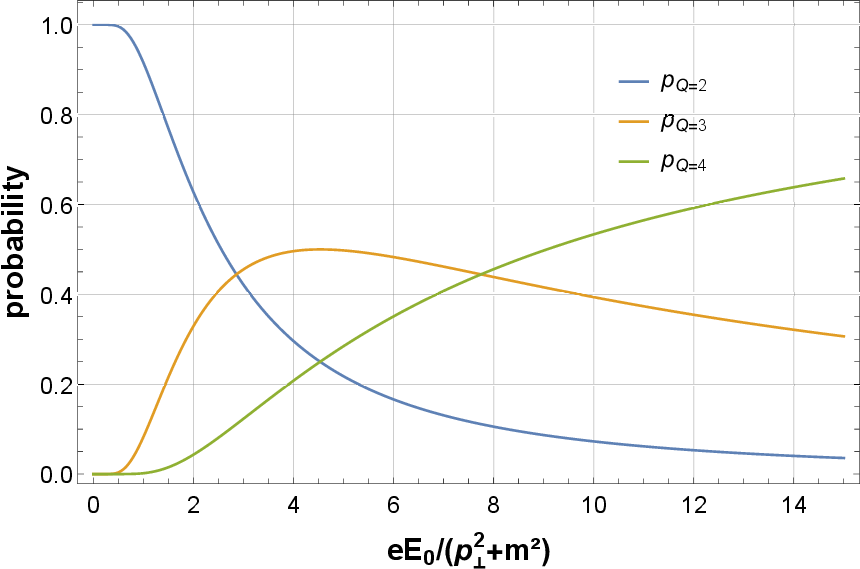}
  \caption{(a) Total symmetry resolved entanglement entropy
    $S=S_{\rm config}+S_{\rm num}$ of $\rho_{\mathbf{p},\mathbf{q}}$ as a
    function of the dimensionless field strength $eE_0/(p_\perp^2+m^2)$,
    for $\varepsilon=1/\sqrt{2},\,1/\sqrt{3},\,1/2,\,1/\sqrt{5},\,1/3$,
    with $p_\perp^2=q_\perp^2$. (b) The corresponding charge sector
    probabilities $p_{Q=2}=|\alpha|^4$, $p_{Q=3}=2|\alpha|^2|\beta|^2$,
    $p_{Q=4}=|\beta|^4$ for the sectors
    $Q=n_{\mathbf{p}}+n_{\mathbf{q}}=2,3,4$; these are independent of
    $\varepsilon$.}
  \label{fig:sree}
\end{figure}

Figure~\ref{fig:sree}(a) shows the total symmetry resolved entanglement entropy of $\rho_{\mathbf{p},\mathbf{q}}$, obtained by resolving the entropy of the bipartition
$(\mathbf{p},\mathbf{q})\,|\,(-\mathbf{p},-\mathbf{q})$ with respect to the
subsystem charge $Q=n_{\mathbf{p}}+n_{\mathbf{q}}$ and summing the
configurational and number contributions,
$S=\sum_Q p_Q S_Q-\sum_Q p_Q\log_2 p_Q$. Since both retained modes are
particles, the physical charge coincides with the total occupation here.
For every $\varepsilon$ the entropy vanishes at zero field, rises steeply,
reaches a maximum near the critical field, and decays slowly thereafter.
The vanishing at $E_0=0$ reflects that this cut measures the entanglement
between the particle modes and the antiparticle modes, which are
unoccupied in the absence of the field. The curves are ordered by the
initial entanglement, $\varepsilon=1/\sqrt{2}$ lying uppermost, but remain
closely bunched: the ordering enters only through the configurational term
$S_{Q=3}=S(\varepsilon)+1$, while the dominant, $\varepsilon$ independent
number entropy $S_{\rm num}$ sets the common shape.

That common shape is explained by Fig.~\ref{fig:sree}(b), which shows the charge sector probabilities $p_Q$. These follow a shifted binomial in the
pair-production probability $|\beta|^2=e^{-\pi\mu}$: $p_{Q=2}=|\alpha|^4$
decreases monotonically from unity, $p_{Q=4}=|\beta|^4$ grows from zero,
and $p_{Q=3}=2|\alpha|^2|\beta|^2$ rises to a maximum of $\tfrac{1}{2}$ at
$|\beta|^2=\tfrac{1}{2}$, i.e.\ at $E_0=\pi(p_\perp^2+m^2)/(e\ln 2)$. As
the field increases, the distribution migrates through the sectors
$Q=2\to3\to4$, tracking the deposition of Schwinger created particles into the modes $\mathbf{p}$ and $\mathbf{q}$. Because only the sector $Q=3$
carries entanglement ($S_{Q=2}=S_{Q=4}=0$, $S_{Q=3}=S(\varepsilon)+1$),
the entropy in Fig.~\ref{fig:sree}(a) inherits its non-monotonic profile
directly from $p_{Q=3}$, peaking at the same critical field. The sector
probabilities are independent of $\varepsilon$, confirming that here the
field controls only the distribution of entanglement over charge sectors,
while the initial state fixes the entanglement carried within the
populated sector.

\subsection{Antiparticle--antiparticle modes}

The antiparticle--antiparticle reduced state $\rho_{-\mathbf{p},-\mathbf{q}}$
follows from $\rho_{\mathbf{p},\mathbf{q}}$ by the exchange
$\alpha\leftrightarrow\beta$, since particle modes are governed by
$|\alpha|^2$ and antiparticle modes by $|\beta|^2$. Resolved by its
physical charge $Q=-(n_{-\mathbf{p}}+n_{-\mathbf{q}})$ (equivalently, by
the antiparticle occupation), its symmetry resolved entanglement entropy
has the same structure as in the particle--particle case, with the sector
probabilities $p_{Q}=|\beta|^4,\,2|\alpha|^2|\beta|^2,\,|\alpha|^4$ and the
same per-sector entropies $0,\,S(\varepsilon)+1,\,0$. The configurational
and number contributions are identical in form, so we do not display them
separately.

The physical content, however, is complementary. Whereas the
particle--particle entanglement is maximal at zero field and is depleted
as $E_0$ grows, the antiparticle--antiparticle entanglement vanishes at
$E_0=0$, there are no antiparticles in the absence of the field and builds up monotonically with increasing $E_0$, since $|\beta|^2$ grows from
zero. The configurational term
$S_{\rm config}=2|\alpha|^2|\beta|^2\,[S(\varepsilon)+1]$ is common to both
and again peaks at the critical field $|\beta|^2=\tfrac{1}{2}$. Thus the
electric field does not simply create or remove resolved entanglement but
transfers it from the particle sector to the antiparticle sector, a
redistribution that mirrors, at the level of individual charge sectors,
the conservation relations obtained for the total quantities.

\subsection{Particle--antiparticle modes}
\begin{figure}[t]
  \centering
  \includegraphics[width=0.48\textwidth]{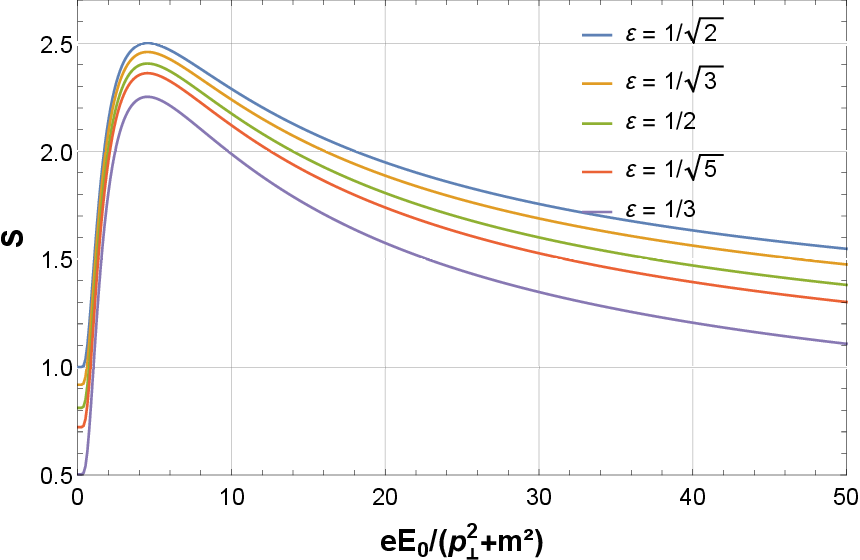}\hfill
  \includegraphics[width=0.48\textwidth]{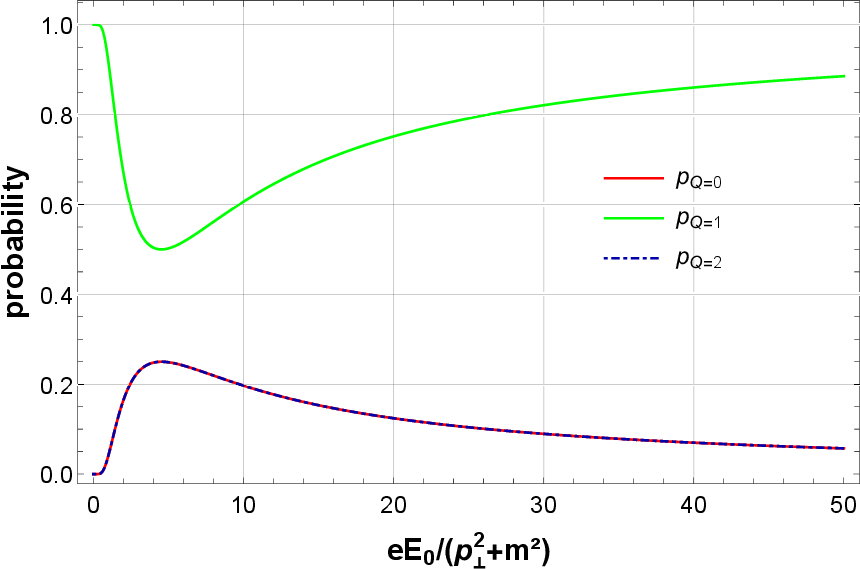}
  \caption{(a) Total symmetry resolved entanglement entropy
    $S=S_{\rm config}+S_{\rm num}$ of $\rho_{\mathbf{p},-\mathbf{q}}$,
    resolved by the physical charge $Q=n_{\mathbf{p}}-n_{-\mathbf{q}}$, as
    a function of the dimensionless field strength $eE_0/(p_\perp^2+m^2)$,
    for $\varepsilon=1/\sqrt{2},\,1/\sqrt{3},\,1/2,\,1/\sqrt{5},\,1/3$,
    with $p_\perp^2=q_\perp^2$. (b) The corresponding charge sector
    probabilities $p_{Q=0}=|\alpha|^2|\beta|^2$,
    $p_{Q=1}=|\alpha|^4+|\beta|^4$, $p_{Q=2}=|\alpha|^2|\beta|^2$ for the
    sectors $Q=0,1,2$; these are independent of $\varepsilon$.}
  \label{fig:sree_pmq}
\end{figure}

Figure~\ref{fig:sree_pmq}(a) shows the total symmetry resolved entanglement entropy of the particle--antiparticle reduced state
$\rho_{\mathbf{p},-\mathbf{q}}$. Since the four-mode state is pure,
$\rho_{\mathbf{p},-\mathbf{q}}$ is the reduced density matrix of the
bipartition $(\mathbf{p},-\mathbf{q})\,|\,(-\mathbf{p},\mathbf{q})$, and
its von Neumann entropy is a genuine entanglement entropy. Because
$\mathbf{p}$ is a particle mode and $-\mathbf{q}$ an antiparticle mode, the
conserved $U(1)$ charge of this subsystem is the physical charge
$Q=n_{\mathbf{p}}-n_{-\mathbf{q}}$, and we resolve with respect to it. The
block decomposition yields three sectors, $Q=0,1,2$, with sector
probabilities
\begin{equation}
  p_{Q=0}=|\alpha|^2|\beta|^2,\quad
  p_{Q=1}=|\alpha|^4+|\beta|^4,\quad
  p_{Q=2}=|\alpha|^2|\beta|^2,
\end{equation}
and per-sector entropies
\begin{equation}
  S_{Q=0}=S_{Q=2}=0,
  \quad
  S_{Q=1}=S(\varepsilon)+H\!\left(\frac{|\alpha|^4}{|\alpha|^4+|\beta|^4}\right),
\end{equation}
where $H(x)=-x\log_2 x-(1-x)\log_2(1-x)$ is the binary entropy, so that
$S=p_{Q=1}\,S_{Q=1}+S_{\rm num}$.

The structure differs qualitatively from the particle--particle case. The
entanglement is now confined to the middle sector $Q=1$; the
extremal sectors $Q=0,2$ are locally pure. Furthermore, $S_{Q=1}$ is
itself field-dependent: it carries a fixed contribution $S(\varepsilon)$
from the initial entanglement plus a field-dependent term
$H(|\alpha|^4/(|\alpha|^4+|\beta|^4))$, which reaches its maximum of one
bit at the critical field $|\beta|^2=\tfrac{1}{2}$, where
$S_{Q=1}=S(\varepsilon)+1$. Thus, in contrast to the particle--particle and
antiparticle--antiparticle pairs, for which the per-sector entropy is
field-independent and all field dependence resides in the sector weights,
the particle--antiparticle entropy acquires field dependence in
both the weights and the entropy of the populated sector. The total
entropy in Fig.~\ref{fig:sree_pmq}(a) rises from zero, peaks near the
critical field, and decays slowly thereafter, ordered by the initial
entanglement with $\varepsilon=1/\sqrt{2}$ uppermost.

Figure~\ref{fig:sree_pmq}(b) shows the sector probabilities. The middle
sector dominates: $p_{Q=1}=|\alpha|^4+|\beta|^4$ falls from unity at zero
field to its minimum $\tfrac{1}{2}$ at the critical field
$|\beta|^2=\tfrac{1}{2}$ and returns towards unity at strong field, while
the two extremal sectors $p_{Q=0}=p_{Q=2}=|\alpha|^2|\beta|^2$ each rise to
$\tfrac{1}{4}$ there. Since only $Q=1$ carries entanglement, the total
entropy reflects the interplay of the weight $p_{Q=1}$ and the
field-dependent sector entropy $S_{Q=1}$, again peaking near the critical
field.

By the combined interchange $\mathbf{p}\leftrightarrow\mathbf{q}$ and
$\varepsilon\leftrightarrow\sqrt{1-\varepsilon^2}$, together with the
invariance $S(\varepsilon)=S(\sqrt{1-\varepsilon^2})$, the total
symmetry resolved entanglement entropy of $\rho_{-\mathbf{p},\mathbf{q}}$
is identical to that of $\rho_{\mathbf{p},-\mathbf{q}}$ for all field
strengths and transverse momenta, and we do not display it separately. The
two particle--antiparticle cross terms are distinguished only at the level
of the logarithmic negativity, whose values,
$N(\rho_{\mathbf{p},-\mathbf{q}})\propto|\alpha_{\mathbf{p}}|^2|\beta_{\mathbf{q}}|^2$
and
$N(\rho_{-\mathbf{p},\mathbf{q}})\propto|\alpha_{\mathbf{q}}|^2|\beta_{\mathbf{p}}|^2$,
differ when $p_\perp^2\neq q_\perp^2$.

\subsection{Vacuum pair}

\begin{figure}[t]
  \centering
  \includegraphics[width=0.48\textwidth]{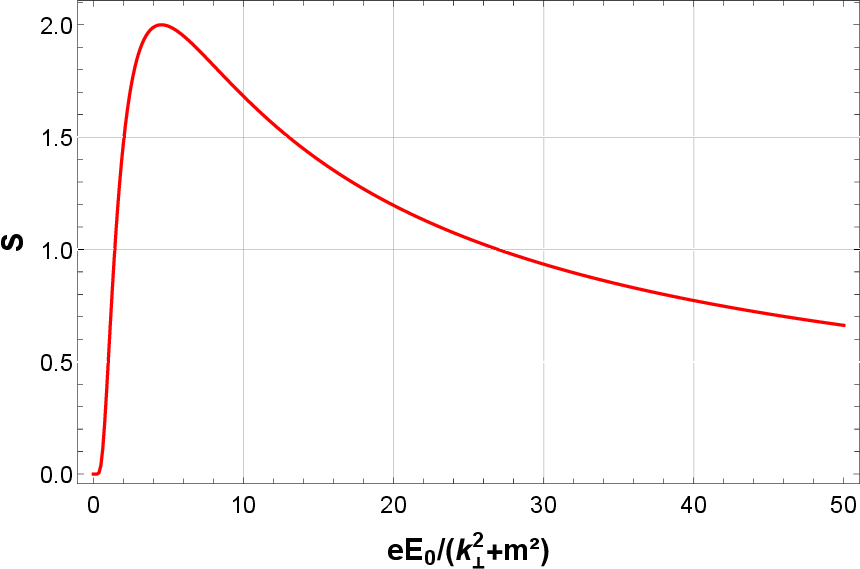}\hfill
  \includegraphics[width=0.48\textwidth]{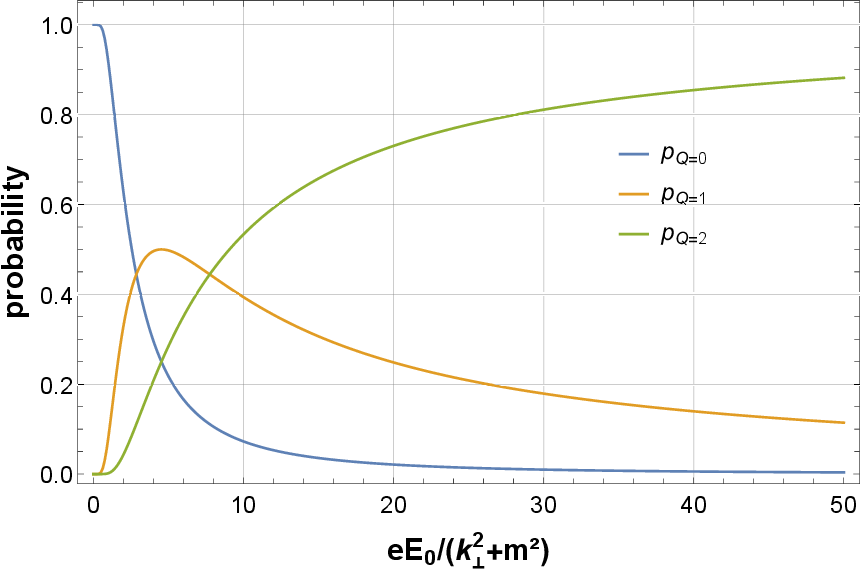}
  \caption{(a) Total symmetry resolved entanglement entropy
    $S=S_{\rm config}+S_{\rm num}$ of the vacuum pair
    $\rho_{\mathbf{k},-\mathbf{k}}$ as a function of the dimensionless
    field strength $eE_0/(k_\perp^2+m^2)$. The curve is independent of
    $\varepsilon$; it vanishes at zero field, decreases towards zero as
    $E_0\to\infty$, and attains its maximum value of $2$ at the critical
    field $|\beta|^2=\tfrac{1}{2}$. (b) The corresponding charge sector
    probabilities $p_{Q=0}=|\alpha|^4$, $p_{Q=1}=2|\alpha|^2|\beta|^2$,
    $p_{Q=2}=|\beta|^4$ for the mode charge $Q=n_{\mathbf{k}}=0,1,2$.}
  \label{fig:sree_vac}
\end{figure}

We now turn to the pair created directly from the vacuum, whose in-vacuum
state is a superposition of out-mode occupations of $\mathbf{k}$ and
$-\mathbf{k}$. In contrast to the previous cases, no pre-existing
entanglement is present: the state contains no parameter $\varepsilon$,
and all of the entanglement is generated by the field itself. Since
$\rho_{\mathbf{k},-\mathbf{k}}$ is a pure two-mode state, the relevant
entanglement is that between the particle mode $\mathbf{k}$ and the
antiparticle mode $-\mathbf{k}$, quantified by the von Neumann entropy of
the reduced state $\rho_{\mathbf{k}}=\mathrm{Tr}_{-\mathbf{k}}\,
\rho_{\mathbf{k},-\mathbf{k}}$. This reduced state is diagonal, with
eigenvalues $|\alpha|^4$, $|\alpha|^2|\beta|^2$, $|\alpha|^2|\beta|^2$, and
$|\beta|^4$, and we resolve its entropy by the mode charge
$Q=n_{\mathbf{k}}$; as $\mathbf{k}$ is a particle mode, this is again the
physical charge.

The resolution yields three sectors, $Q=0,1,2$, with per sector entropies
$S_{Q=0}=S_{Q=2}=0$ and $S_{Q=1}=1$, so that
\begin{equation}
  S_{\rm config}=2|\alpha|^2|\beta|^2,
  \qquad
  S=S_{\rm config}+S_{\rm num}.
\end{equation}
Only the singly occupied sector carries entanglement, and its value is
fixed at one bit; the entire field dependence enters through the sector
weights. Figure~\ref{fig:sree_vac}(a) shows the resulting total entropy:
it grows from zero, reaches a maximum at the critical field
$E_0=\pi(k_\perp^2+m^2)/(e\ln 2)$, and decreases slowly thereafter. The
maximum equals exactly $2$, corresponding to the four out-mode
configurations of $\rho_{\mathbf{k}}$ becoming equiprobable at
$|\alpha|^2=|\beta|^2=\tfrac{1}{2}$. This non-monotonic profile parallels
that of the pair-creation negativity and reflects that both are controlled
solely by the pair-production probability $|\beta|^2=e^{-\pi\mu}$.

Figure~\ref{fig:sree_vac}(b) shows the charge sector probabilities, a
binomial distribution in $|\beta|^2$: $p_{Q=0}=|\alpha|^4$ decreases
monotonically from unity, $p_{Q=2}=|\beta|^4$ grows from zero, and
$p_{Q=1}=2|\alpha|^2|\beta|^2$ peaks at $\tfrac{1}{2}$ at the critical
field. As the field increases, the distribution migrates from the empty
sector $Q=0$ through the singly-occupied sector to the fully-occupied
sector $Q=2$, tracking the progressive conversion of the vacuum into
particle--antiparticle pairs. Because only the intermediate sector carries
entanglement, the total entropy in Fig.~\ref{fig:sree_vac}(a) follows the
weight $p_{Q=1}$ of that sector, peaking where pair creation is most
balanced.

\begin{figure}[t]
  \centering
  \includegraphics[width=0.5\textwidth]{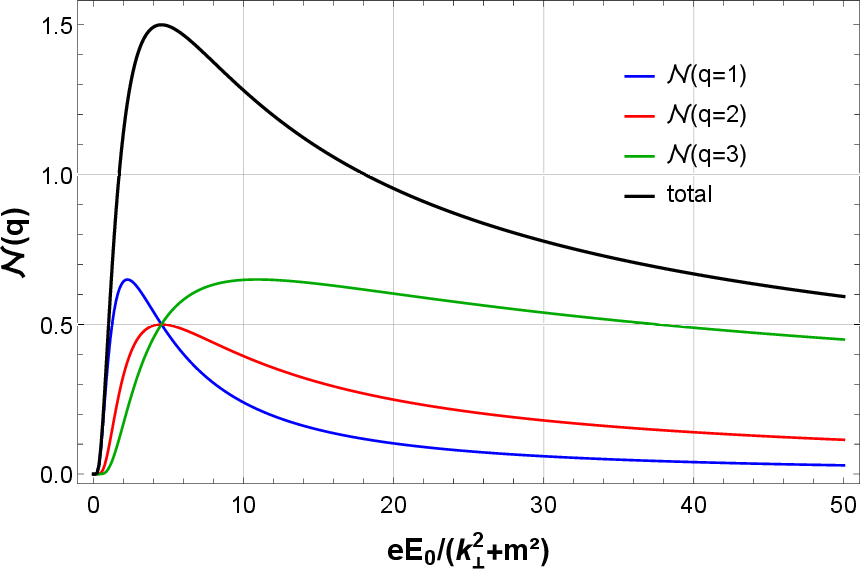}
  \caption{Symmetry resolved negativity of the vacuum pair
    $\rho_{\mathbf{k},-\mathbf{k}}$ as a function of the dimensionless
    field strength $eE_0/(k_\perp^2+m^2)$. The coloured curves are the
    resolved negativities of the imbalance sectors,
    $\mathcal{N}(q{=}1)=2|\alpha|^3|\beta|$,
    $\mathcal{N}(q{=}2)=2|\alpha|^2|\beta|^2$, and
    $\mathcal{N}(q{=}3)=2|\alpha||\beta|^3$, with the imbalance charge
    $q=Q_A-Q_B=n_{\mathbf{k}}+\bar{n}_{-\mathbf{k}}$; the black curve is
    their sum, the total negativity $\mathcal{N}$. The empty sectors $q=0$
    and $q=4$ carry no negativity.}
  \label{fig:neg_vac}
\end{figure}

Figure~\ref{fig:neg_vac} shows the symmetry resolved negativity of the
vacuum pair. Being mixed at the level of a single mode, the appropriate
entanglement measure is the logarithmic negativity, which we resolve with
respect to the charge imbalance
$q=Q_A-Q_B=n_{\mathbf{k}}+\bar{n}_{-\mathbf{k}}$ that commutes with the
partial transpose $\rho_{\mathbf{k},-\mathbf{k}}^{T_A}$; here
$Q_A=n_{\mathbf{k}}$ is the particle charge and $Q_B=-\bar{n}_{-\mathbf{k}}$
the antiparticle charge, so the imbalance is the total number of
excitations. The partial transpose block-diagonalises into the sectors
$q=0,\dots,4$, and summing the moduli of its negative eigenvalues within
each block gives
\begin{equation}
  \begin{aligned}
  \mathcal{N}(q{=}1)&=2|\alpha|^3|\beta|,\\
  \mathcal{N}(q{=}2)&=2|\alpha|^2|\beta|^2,\\
  \mathcal{N}(q{=}3)&=2|\alpha||\beta|^3,
  \end{aligned}
  \label{eq:Nq_vac}
\end{equation}
with $\mathcal{N}(q{=}0)=\mathcal{N}(q{=}4)=0$ and
$\sum_q \mathcal{N}(q)=\mathcal{N}$. The total reproduces the unresolved
result, $\log_2(1+2\mathcal{N})=4\log_2(|\alpha|+|\beta|)$. Because $\rho_{\mathbf{k},-\mathbf{k}}$ is a pure state, its logarithmic
negativity is fixed by the Schmidt spectrum alone and is therefore
independent of the transpose convention: the ordinary and the fermionic
partial transpose yield the same trace
norm, so the total $\mathcal{N}=2(|\alpha|+|\beta|)^{2}-1$ and hence
$\log_{2}(1+2\mathcal{N})=4\log_{2}(|\alpha|+|\beta|)$ hold under either
definition. The headline result for the vacuum pair thus does not rely on
the bosonic-style partial transpose used above. We show in Appendix~A that this imbalance resolved decomposition, and not only the total, is unchanged under the fermionic partial
transpose~\cite{Shapourian2017,Eisler2015}.

In contrast to the reduced states of the initially entangled pair, whose
negativity is confined to a single charge sector, here the negativity is
distributed over three sectors. Each contribution is non-monotonic and
peaks at a different field: writing $|\beta|^2=e^{-\pi\mu}$, the maxima of
$\mathcal{N}(q{=}1)$, $\mathcal{N}(q{=}2)$, and $\mathcal{N}(q{=}3)$ occur
at $|\beta|^2=\tfrac{1}{4},\tfrac{1}{2},\tfrac{3}{4}$, respectively. As the
field increases, the dominant contribution therefore migrates through the
ladder $q=1\to2\to3$: at weak field the negativity resides mainly in the
single-excitation sector, at strong field in the triply-occupied sector,
and only at the critical field $|\beta|^2=\tfrac{1}{2}$ do the three
sectors contribute equally. The electric field thus does not merely raise
or lower the total pair entanglement, it transports it between charge
sectors, with the crossover locked to the pair-production probability
$|\beta|^2$. This field tuned departure from equal sector contributions
contrasts with the exact equipartition of negativity found for massless
free fermions, which is recovered here only at the single balanced point.

\subsection{Partner matrices $\rho_{\mathbf{p},-\mathbf{p}}$ and
$\rho_{\mathbf{q},-\mathbf{q}}$}

For the same momentum partner matrices we consider the reduced state
$\rho_{\mathbf{p},-\mathbf{p}}=\mathrm{Tr}_{\mathbf{q},-\mathbf{q}}
|\Phi\rangle\langle\Phi|$, obtained by tracing out the $\mathbf{q}$-sector.
Its entanglement entropy is that of the bipartition
$(\mathbf{p},-\mathbf{p})\,|\,(\mathbf{q},-\mathbf{q})$. A direct
evaluation shows that $\rho_{\mathbf{p},-\mathbf{p}}$ is a rank-two mixture
of two orthonormal states with weights $\varepsilon^2$ and
$1-\varepsilon^2$, so its eigenvalues are $\varepsilon^2$ and
$1-\varepsilon^2$ and
\begin{equation}
  S(\rho_{\mathbf{p},-\mathbf{p}})=S(\varepsilon),
\end{equation}
independent of the electric field. The physical charge of this subsystem
is $Q=n_{\mathbf{p}}-n_{-\mathbf{p}}$, and $\rho_{\mathbf{p},-\mathbf{p}}$
lies entirely within the single sector $Q=1$ (it does not commute with the
total occupation $n_{\mathbf{p}}+n_{-\mathbf{p}}$). The symmetry
resolution is therefore trivial: a single sector of unit weight carrying
the full entropy $S(\varepsilon)$, with vanishing number entropy.

This field independence has a simple origin. Schwinger pair creation acts
only within each momentum sector depositing a particle in $\mathbf{p}$
together with an antiparticle in $-\mathbf{p}$, and likewise for
$\mathbf{q}$---and thus conserves $n_{\mathbf{p}}-n_{-\mathbf{p}}$. It never
generates entanglement across the
$(\mathbf{p},-\mathbf{p})\,|\,(\mathbf{q},-\mathbf{q})$ cut, so the
entanglement of that cut remains fixed at the preexisting value
$S(\varepsilon)$ for all field strengths. The complementary partner matrix
$\rho_{\mathbf{q},-\mathbf{q}}$ is the reduced state of the complementary
cut of the same pure state and follows by
$\mathbf{p}\leftrightarrow\mathbf{q}$; its symmetry-resolved entanglement
entropy is likewise $S(\varepsilon)$, field-independent and confined to the
sector $Q=n_{\mathbf{q}}-n_{-\mathbf{q}}=1$. We therefore do not display
these cases separately. The nontrivial, field-dependent content of the
partner matrices resides instead in their logarithmic negativity.

Resolving it with respect to the imbalance charge under the ordinary partial
transpose yields only a single sector: the negative eigenvalues of
$\rho^{T_{A}}_{\mathbf{p},-\mathbf{p}}$ all lie in the imbalance sector
$q=n_{\mathbf{p}}+n_{-\mathbf{p}}=2$, so the resolved negativity coincides
with the total and carries no additional charge structure, as for the
particle--antiparticle cross terms. Unlike the vacuum pair, however, these
are genuinely mixed states, for which the ordinary and fermionic partial
transpose need not agree~\cite{Shapourian2017,Eisler2015}; the pure state
argument of Appendix~A therefore does not apply, and the single sector
structure quoted here should be understood as the ordinary transpose result.
A complete charge imbalance resolution of the mixed negativities based on the
fermionic partial transpose is left for future work. We therefore do not
display these cases separately.
\section{Conclusion}

In this work we have refined the study of pairwise mode entanglement in
the Schwinger effect by resolving it with respect to the underlying $U(1)$
electric charge. Whereas the total entanglement measures give a single
number at each field strength, symmetry resolution decomposes that number
into contributions from individual charge sectors, revealing where in
charge space the entanglement resides and how the electric field moves it.
We chose the measure according to the state: the symmetry resolved
entanglement entropy, resolved by the physical charge $Q$, for the
pure bipartition reductions, and the charge imbalance resolved negativity,
resolved by the imbalance charge $q$, for the mixed particle--antiparticle
pair.

Our central findings are as follows. For $\rho_{\mathbf{p},\mathbf{q}}$ the
resolved entropy is carried by charge sectors
$Q=n_{\mathbf{p}}+n_{\mathbf{q}}=2,3,4$ whose internal entropies are
field-independent, so that the entire field dependence enters through the
sector probabilities, a shifted binomial in the pair-production
probability $|\beta|^2$, peaked at the critical field
$|\beta|^2=\tfrac{1}{2}$. The antiparticle--antiparticle matrix
$\rho_{-\mathbf{p},-\mathbf{q}}$ is the $\alpha\leftrightarrow\beta$ image
of this case, with complementary field dependence, so that the field
transfers resolved entanglement from the particle to the antiparticle
sector. The particle--antiparticle matrix $\rho_{\mathbf{p},-\mathbf{q}}$,
resolved by its physical charge $Q=n_{\mathbf{p}}-n_{-\mathbf{q}}$, behaves
differently: the entanglement is confined to the single middle sector
$Q=1$, whose entropy is itself field-dependent,
$S_{Q=1}=S(\varepsilon)+H(|\alpha|^4/(|\alpha|^4+|\beta|^4))$, so that field
dependence resides in both the sector weight and the sector entropy. The
vacuum pair $\rho_{\mathbf{k},-\mathbf{k}}$ is the richest case: its
resolved negativity is distributed over three imbalance sectors,
$\mathcal{N}(q)=2|\alpha|^{4-q}|\beta|^{q}$, whose maxima occur at
$|\beta|^2=\tfrac{1}{4},\tfrac{1}{2},\tfrac{3}{4}$, so that the field
transports the negativity up the charge ladder $q=1\to2\to3$ and the
sectors contribute equally only at the critical field. Finally, the
same momentum partner matrices have entanglement frozen at $S(\varepsilon)$
across the $(\mathbf{p},-\mathbf{p})\,|\,(\mathbf{q},-\mathbf{q})$ cut,
confined to a single charge sector, because pair creation conserves
$n_{\mathbf{p}}-n_{-\mathbf{p}}$. Overall, the field does not simply create
or destroy pairwise entanglement but redistributes it across charge sectors
and mode pairings, with all field dependence traceable to the single
quantity $|\beta|^2$; symmetry resolution thereby sharpens the conservation
relations of the total measures into a sector resolved picture.

Several directions remain open. The present analysis was carried out for a
constant field; extending the sector decomposition to the Sauter pulsed
field would introduce the pulse width as a second control parameter and
test how robustly the charge sector picture survives a time-dependent
background. The symmetry resolved negativity of the same-momentum partner
modes calls for a careful treatment based on the fermionic partial
transpose, which we have left aside. It would also be worthwhile to relax
the symmetric momentum assumption, and to explore whether the
sector-resolved and equipartition features found here carry over to the
closely related particle-creation processes, the Unruh and Hawking effects
and cosmological particle production to which the Schwinger effect is
formally analogous. More broadly, our results add to the growing use of
charge resolved entanglement as a diagnostic in quantum field
theory~\cite{Goldstein2018,Cornfeld2018,Bonsignori2019,Florio2024}, and
suggest it as a fine grained probe of the Schwinger mechanism beyond the
information contained in particle-number distributions.

\section{Acknowledgment}
Author Amaria Javed acknowledges the support by Tamkeen under the NYU (New York University, Abu Dhabi) Research Institute  grant CG008.	
\section*{Appendix A: Fermionic partial transpose of the vacuum pair}
\label{app:fpt}

Here we show that the imbalance resolved negativities of the vacuum pair,
Eq.~(21), are unchanged when the ordinary partial transpose is replaced by
the fermionic partial transpose of Refs.~\cite{Shapourian2017,Eisler2015},
so that the result of Sec.~3.4 is not an artefact of the transpose
convention.

\paragraph{Factorized form.}
The Bogoliubov transformation~(3)--(4) pairs each particle mode $(\sigma,\mathbf{k})$
with the antiparticle mode $(-\sigma,-\mathbf{k})$. The in-vacuum~(7) therefore
factorizes into two independent two-mode channels,
\begin{equation}
\begin{aligned}
|0_{\mathbf{k}},0_{-\mathbf{k}}\rangle_{\rm in}
= |\chi\rangle_{\uparrow\mathbf{k},\,\downarrow(-\mathbf{k})}
\otimes
|\chi\rangle_{\downarrow\mathbf{k},\,\uparrow(-\mathbf{k})},\\
|\chi\rangle = |\alpha|\,|00\rangle + |\beta|\,|11\rangle ,
\label{eq:A-factorized}
\end{aligned}
\end{equation}
where $|00\rangle$ and $|11\rangle$ are the empty and doubly occupied states
of a channel and local phases have been absorbed, as they do not affect the
moduli below. We label the four fermionic modes
$1=\,\uparrow\!\mathbf{k}$, $2=\,\downarrow\!\mathbf{k}$ (both in $A$) and
$3=\,\downarrow\!(-\mathbf{k})$, $4=\,\uparrow\!(-\mathbf{k})$ (both in $B$),
so that channel~1 joins $(1,3)$ and channel~2 joins $(2,4)$. Writing
$x\equiv|\alpha|$, $y\equiv|\beta|$ with $x^{2}+y^{2}=1$ and occupations as
$|n_{1}n_{2};n_{3}n_{4}\rangle$,
\begin{equation}
|\Psi\rangle
= x^{2}\,|00;00\rangle
+ xy\,|10;10\rangle
+ xy\,|01;01\rangle
+ y^{2}\,|11;11\rangle .
\label{eq:A-state}
\end{equation}

\paragraph{Ordinary partial transpose.}
Transposing $\rho=|\Psi\rangle\langle\Psi|$ on $A$ maps
$|a_{i}b_{i}\rangle\langle a_{j}b_{j}| \mapsto |a_{j}b_{i}\rangle\langle a_{i}b_{j}|$.
The transposed matrix commutes with the imbalance
$\hat q=\hat Q_{A}-\hat Q_{B}=n_{\mathbf{k}}+\bar n_{-\mathbf{k}}$ and
block-diagonalizes into the sectors $q=0,\dots,4$. Crucially, although the
original state populates only total occupations $0,2,4$, the transpose
generates matrix elements in the odd sectors $q=1,3$.

The $q=1$ block acts on $\{|10;00\rangle,|01;00\rangle,|00;10\rangle,
|00;01\rangle\}$ and contains no diagonal (probability) entries; it is a
direct sum of two antidiagonal $2\times2$ blocks,
\begin{equation}
\begin{pmatrix} 0 & x^{3}y \\ x^{3}y & 0 \end{pmatrix}
\ \oplus\
\begin{pmatrix} 0 & x^{3}y \\ x^{3}y & 0 \end{pmatrix},
\qquad \text{eigenvalues } \pm x^{3}y \ \text{(twice)} .
\label{eq:A-q1}
\end{equation}
Summing the moduli of the negative eigenvalues gives
$\mathcal{N}(q{=}1)=2x^{3}y=2|\alpha|^{3}|\beta|$. By the interchange
$x\leftrightarrow y$ the $q=3$ block yields
$\mathcal{N}(q{=}3)=2xy^{3}=2|\alpha||\beta|^{3}$. The $q=2$ block carries
two diagonal entries $x^{2}y^{2}$ together with two antidiagonal $2\times2$
blocks $\bigl(\begin{smallmatrix}0&x^{2}y^{2}\\x^{2}y^{2}&0\end{smallmatrix}\bigr)$,
whose two negative eigenvalues sum to
$\mathcal{N}(q{=}2)=2x^{2}y^{2}=2|\alpha|^{2}|\beta|^{2}$. The empty sectors
$q=0,4$ carry none. This reproduces Eq.~(21), and the sum
$\sum_{q}\mathcal{N}(q)=2xy\,(x^{2}+xy+y^{2})$ coincides with the total
$\mathcal{N}=\tfrac12[(x+y)^{4}-1]$, i.e.\ $\log_{2}(1+2\mathcal{N})=
4\log_{2}(|\alpha|+|\beta|)$.

\paragraph{Fermionic partial transpose.}
The fermionic partial transpose $\rho^{R_{A}}$ differs from the ordinary one
only by unimodular phases $i^{\,\phi(n_{A},\bar n_{A})}$ that multiply each
matrix element according to its $A$-occupations~\cite{Shapourian2017,Eisler2015};
being diagonal in the occupation numbers, these phases neither connect states
of different $q$ nor generate new nonzero elements, so the block structure
above is preserved. The odd sectors $q=1,3$ consist entirely of antidiagonal
$2\times2$ blocks with no diagonal part; multiplying their off-diagonal
entries by phases leaves the singular values $|x^{3}y|$ and $|xy^{3}|$ and
hence $\mathcal{N}(q{=}1)$ and $\mathcal{N}(q{=}3)$ invariant. Since the state
is pure, the \emph{total} negativity is transpose independent, both the
ordinary and the fermionic prescription giving
$\log_{2}(1+2\mathcal{N})=4\log_{2}(|\alpha|+|\beta|)$. The remaining sector
is then fixed by subtraction,
$\mathcal{N}(q{=}2)=\mathcal{N}-\mathcal{N}(q{=}1)-\mathcal{N}(q{=}3)
=2|\alpha|^{2}|\beta|^{2}$. Equation~(21) therefore holds under the fermionic
partial transpose as well, and the transport of negativity up the ladder
$q=1\to2\to3$ is a genuine, convention-independent feature of the vacuum pair.

\end{document}